\documentclass[preprint,times,11pt]{elsarticle}

\usepackage{enumerate}
\usepackage{lineno}
\usepackage{amsmath}
\usepackage{amssymb}
\usepackage{amsfonts}
\usepackage{miller}
\usepackage{graphicx}
\usepackage{textcomp}
\usepackage{nicefrac}
\usepackage{siunitx}
\usepackage[font=footnotesize,labelfont=bf]{subcaption}
\usepackage{boldline}
\usepackage{overpic}      
\usepackage{stackengine}  %
\usepackage{placeins} 
\usepackage{datetime}
\usepackage{hyperref}

\def\degree{$^{\circ}$}

\usepackage{geometry}
\newcommand{\mrk}[1]{#1}

\begin{document}

\title{Tetragonality mapping of martensite in a high-carbon steel by EBSD}

\author[label2,label6]{Gert Nolze}
\address[label2]{Federal Institute for Materials, Research and Testing (BAM), Unter den Eichen 87, 12205 Berlin, Germany}
\ead{Gert.Nolze@bam.de}
\address[label6]{TU Bergakademie Freiberg, Institute for Mineralogy, Brennhausgasse 14, 09596 Freiberg, Germany}

\author[label1,label5]{Aimo Winkelmann}
\address[label1]{Academic Centre for Materials and Nanotechnology, AGH University of Science and Technology, al.\@ A. Mickiewicza 30, 30-059 Krakow, Poland}

\author[label1]{Grzegorz Cios}

\address[label5]{Department of Physics, SUPA, University of Strathclyde, Glasgow G4 0NG, United Kingdom}

\author[label1]{Tomasz Tokarski}

\begin{abstract}
The locally varying tetragonality in martensite grains of a high-carbon steel (1.2\,mass percent C) was resolved by electron backscatter diffraction (EBSD) with a spatial resolution in the order of 100\,nm. 
Compared to spatially integrating X-ray diffraction, which yielded an average tetragonality of $c/a=1.05$, the EBSD measurements in the scanning electron microscope allowed to image a local variation of the lattice parameter ratio $c/a$ in the range of $1.02 \leq c/a \leq 1.07$.
The local variation of tetragonality is confirmed by two different EBSD data analysis approaches based on the fitting of simulated to experimental EBSD patterns. The resulting EBSD-based tetragonality maps are pointing to a complex interaction of carbon concentration and local lattice distortions during the formation process of martensitic structures.

\mbox{\small{\today \, \currenttime}}

\end{abstract}


\maketitle


\section{Introduction} \label{sec:Intro}

The mechanical properties of steels are governed by the interplay of the chemical composition and crystallographic ordering in the microstructure of these materials \cite{bhadeshia2017steels}. 
In this respect, an extraordinary role is played by the formation of martensite, which is understood as a microstructure that is formed by a collective shear of atoms during a diffusionless and athermal solid state transformation from an austenitic parent phase \cite{Nishijama:78,bhattacharya2003microstructure,lobodyuk2005pu,kelly2012ptis}. 
Crystallographically, the martensite formed in this way can be interpreted as a slightly tetragonal distorted ferrite whose $c/a$ ratio varies with carbon content \cite{Kurdjumov:76}.
Chemically, the observed tetragonality of the martensite is assumed to be caused by the carbon in solid solution, with typical concentrations below approximately 10 atomic percent. 

Spatially, the resulting martensite grains have typically sub-micron dimensions \cite{Krauss1999} and are commonly heavily distorted, which is why a locally resolved characterization of martensite is in general challenging.
\mrk{This also explains why, even after decades of investigations, relevant questions about the crystallographic nature of the martensite distortion \cite{lobodyuk2018mmta}, the spatial variation of the tetragonality \cite{christian1992mt,han2001pma}, the influence of local strains on the martensite lattice \cite{mirzaev2016msf,chirkov2016pomm,maugis2020jped}, as well as the effects of aging in Fe-C alloys \cite{genderen1997mmta} are still not conclusively answered and are subject to ongoing debate.}
The majority of previous $c/a$ measurements on martensite are based on powder X-ray diffraction (XRD) experiments performed on polycrystalline bulk material. These results therefore only reflect the lattice parameters averaged over the probed volume \cite{cheng1990smet}.
\mrk{
A comprehensive discussion of the challenges of analyzing the tetragonality of martensite in steels by XRD measurements can be found in \cite{tanaka2020amat}. }
Spatially resolved lattice parameter characterization by XRD, however, would be extremely difficult. A complicating factor is that the observed reflex broadening in x-ray diffractograms of martensite can be interpreted in different ways \cite{Kurdjumov:76,stephens1999jac,leineweber2004jac}. 
\mrk
{It can be attributed to an increase in defect density or to a reduced grain size after transformation \cite{Satdarova:16,Wiessner:14}.
Another interpretation possibility are locally varying lattice parameters \cite{Kurdjumov:76,leineweber2004jac}, e.g.\@ caused by inhomogeneities in the element and/or defect distribution. It is challenging to separate these influences by X-ray diffraction.}

In this context, electron backscatter diffraction (EBSD) is an attractive technique because it offers the possibility to determine the local crystallography of a microstructure with a resolution in the submicrometer range in a scanning electron microscope (SEM) \cite{schwartz2009ebsd2}. 
Several studies have been recently presented for the EBSD-based analysis of tetragonality in martensite grains \cite{Tanaka:18,Winkelmann:18,tanaka2020amat}. These investigations involved the matching of experimental, background corrected EBSD patterns to Kikuchi pattern simulations for a fixed set of tetragonalities \cite{Tanaka:18}, or the consistent deformation of a fixed reference simulation for a variable fit to local $c/a$ ratios and general strain states \cite{Winkelmann:18}. 
\mrk{In the study of Tanaka \textit{et al.} \cite{tanaka2020amat}, a systematic study of the $c/a$ lattice parameter ratio is presented which compares results of spatially averaging XRD measurements with spatially resolving EBSD pattern matching for a limited number of representative map points which were selected on steel samples containing different average amounts of carbon. At the different sample positions, these EBSD measurements showed a variation of the $c/a$ ratio which were significantly different from the averaged XRD value for the respective sample.}

\mrk{The aim of the present study is to demonstrate the possibility of a detailed mapping of the local variation of the tetragonality in martensite with a spatial resolution in the order of 100\,nm, including a comparison to integral X-ray diffraction measurements. We demonstrate that a useful range of local $c/a$ ratios can be resolved even with EBSD patterns having conventional image resolutions of $160\times 115$ pixels ($\approx20$k versus $\approx900$k pixels in \cite{tanaka2020amat}, i.e.\@ only 2\% of the high-resolution image size). 
For the analysis, the projective pattern matching approach presented in \cite{Winkelmann:18} is applied which enables the characterization of large-area EBSD maps, i.e. including a significant number of martensite grains.
The improved spatial resolution, statistical relevance and reliability of the characterization of local lattice parameter variations by EBSD are expected to contribute to an improved understanding of the microscopic formation mechanisms in martensitic structures.}

\section{Experimental Details}
\label{sec: Experimental}

\subsection{Material} \label{sec:Exp-Material}

The investigated material is a high-carbon model steel called 1.2\,C in relation to its carbon content. 
The chemical composition as measured by Glow Discharge Optical Emission Spectroscopy (GD-OES) is given in Table\,\ref{tab:Composition}.
\newcolumntype{C}[1]{>{\centering\arraybackslash}p{#1}}
\renewcommand{\arraystretch}{1.35}
\begin{table}[ht]
\centering
\caption{\label{tab:Composition}Minor element concentrations in m\% for 1.2\,C steel.
(GD-OES).}
\vspace{-1.5ex}
\begin{tabular}{C{1cm}C{1cm}C{1cm}C{1cm}C{1cm}C{1cm}} 
 C & Mn & Si & Cu & P & S \\ \hline
 1.21 & 0.57 & 0.09 & 0.02 & 0.0095 & $0.0057$ \\ 
\end{tabular}
\end{table}

A sample of $15\times 15 \times 3$\,mm$^3$ was cut from a sheet to guarantee high cooling rates after solution heat treatment. After soaking the sample for 30\,min at 1000\degree C to solute cementite in austenite, it was quenched in agitated water. This is above the A$_\textrm{cm}$ or S-E line in the Fe--Fe$_{3}$C phase diagram \cite{krauss:05}. 

The sample was halved parallel to the large surface plane by wire electric discharge machining. On the cut surface a standard metallographic preparation was applied starting with grinding from 240 to 2000 grit SiC papers. The surface was then polished subsequently by 3\,\textmu m and 1\,\textmu m diamond suspensions. Finally, a manual surface finishing with 0.05 \textmu m colloidal silica for 10 min was applied to produce a surface quality free from any visible preparation artifacts. 
The central area of the sample  was chosen to avoid any decarburized zones within the analysis region.

\subsection{XRD} \label{sec:Exp-XRD}

The lattice parameters were derived from an X-ray diffractogram collected in an Empyrean diffractometer (PANalytical). It was equipped with a parabolic mirror as primary beam optics. In order to prevent any Fe-fluorescence,  Co-radiation was applied.  
The diffractogram was acquired in 3\,h and 40\,min for $2\theta=15-130$\degree{} using a step size of $\approx 0.0263$\,\degree. 
The lattice parameters $c$ and $a$ were optimized using a least-square refinement available in HighScore Plus \cite{highscore}. 

For the correlation between carbon concentration $x$ (in m\%) and mean tetragonal distortion $(c/a)_m$ in martensite, we use the following approximate relationship determined from X-ray diffraction studies \cite{Roberts:53,cheng1990smet}:

\begin{align}\nonumber
c_m \textrm{[\AA]}& = 2.861 + 0.116\cdot x \\ \label{equ:c-a}
a_m \textrm{[\AA]}& = 2.861 - 0.013\cdot x \\ \nonumber
(\nicefrac{c}{a})_m & = 1.000 + 0.045\cdot x
\end{align}

with the mean lattice parameters $a_m$ and $c_m$.

\subsection{EBSD} \label{sec:Exp-EBSD}
Diffraction patterns with an image resolution of \mbox{$160\times 115$} pixels were recorded with an $e^{-}$Flash${^\textrm{HR}}$ EBSD detector (Bruker Nano) mounted on a field-emission scanning electron microscope LEO 1530VP (Zeiss).
At an acceleration voltage of 20 kV and a beam current of $\approx 12$\,nA (120\,\textmu m-aperture and high current mode) the typical acquisition time in high vacuum was 20\,ms per pattern which represented approximately 90\% of the saturation limit of the camera. The step size used was 94\,nm.
No additional pattern averaging was applied.
All raw EBSD patterns were stored together with the phase-specific crystal orientations after the measurement was completed.

\subsubsection{Pattern matching approach}

In order to derive crystallographic information from the Kikuchi patterns measured by EBSD, we use a best-fit optimization procedure which maximizes the image similarity between an experimental Kikuchi pattern and a simulated model pattern \cite{Nolze:15c,Winkelmann:16a,Nolze:17,Nolze:18,winkelmann2020jm,nolze2020iop}.
As the image similarity metric for a comparison of two Kikuchi patterns, we used the normalized cross-correlation coefficient (NCC) $r<1$ \cite{kirkland2010}.
We applied the Nelder-Mead approach \cite{neldermead1965tcj} for the maximization of the NCC between the experimental and the simulated patterns. 

In short, the pattern matching analysis of a full EBSD map proceeds in the following steps: (1) calibration of the scan geometry, which results in a consistent map of source point positions relative to the EBSD screen \cite{winkelmann2020mat}, (2) determination of the best-fit orientation for a specific phase (e.g. showing a specific $c/a$ ratio), including testing for orientational pseudosymmetry \cite{Winkelmann:18}, and (3) assignment of the phase based on the relative values of the pattern similarity for the best-fit orientation.

\paragraph{Diffraction geometry}

For the maps considered in this paper, we have determined the projection centers (PC) for the complete map by fitting both the PC position and the orientation (in total 6 fit parameters) using the patterns of the retained cubic austenite in the sample as calibration phase. 
After calibration, the projection center is assumed to be fixed for each point in the map when fitting the orientation and tetragonality of each candidate phase.
The applied calibration procedure is described in \cite{friedrich2018um} and \cite{winkelmann2020mat}. 
The precision of projection center determination using various pattern matching approaches is discussed in \cite{tanaka2019um}. 

\paragraph{Orientation refinement and Phase assignment}

The individual Kikuchi patterns of the map can be tested against simulations for the orientations of each candidate phase \cite{winkelmann2020jm}.  
The highest, best-matching NCC $r$ value defines phase and orientation, which is assigned to the respective measuring point on the map. 
The NCC values $r<0.2$ indicate unreliable solutions and were not considered in the analysis.

As starting values for the orientation refinement we used the orientation descriptions provided by the EBSD system software. 
These orientation data typically do not deviate from the actual orientation by more than 2\degree{} \cite{Krieger-Lassen:96,Ram:15}. 
If a particular pattern could not be correctly solved by the EBSD system software, the orientations of neighboring points are also tested as initial estimates.
If this does not provide reliable starting values, the template matching described in \cite{Chen:15} or \cite{Britton:18a} can be used to find a suitable initial orientation.  
The initial orientations described by three Euler angles are then varied and optimized assuming a locally fixed PC.
To prevent systematic errors of the EBSD system software from affecting the initial values, the initial orientation is also randomly changed by 1\degree. The refinement process is then started with these Euler angles.

For the orientation optimization of the pseudo-cubic martensite, all possible orientation variants must be checked, assuming that $c$ can point in the direction of any of the initial cubic $a$ axes.
All pseudo-symmetric alignments of $c$ are then optimized independently for each pattern. The solution with the best $r$ value is then selected as the valid orientation description.

\section{Results and discussion} \label{sec:Results}

\subsection{Tetragonality of martensite by XRD}

\begin{figure}[htb!]
\centering
\includegraphics[width=0.6\columnwidth]{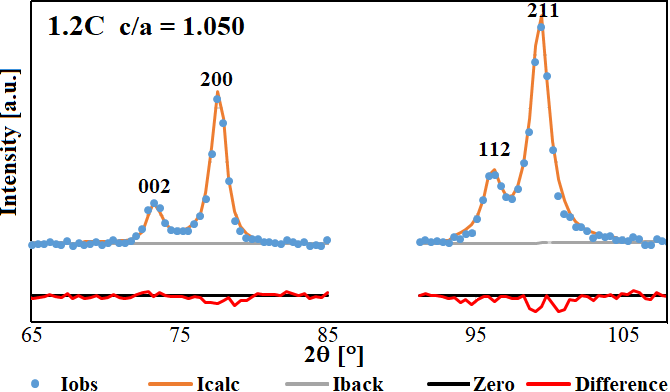}
\caption{ X-ray diffractogram of 1.2\,C. 
The red graph displays the difference plot between experimental (blue dots) and simulated intensity (orange line). Gray and black line denote the fitted background and zero level.}
\label{fig:XRD-Diff}
\end{figure}

The standard technique for determining lattice parameters is XRD. For the steel under investigation the $2\theta$ range between 45\degree{} and 130\degree{} was measured. 
Because of the unknown crystallographic texture \mrk{and the small amount of available reflections}, for the refinement of the experimental signal a LeBail fit \cite{Lebail_2005} was used. This means that all peak intensities are adjusted without the typical correlation to the crystal structure amplitude. \mrk{Thus the fit was optimized not only by varying the lattice parameters but also by assuming uncorrelated reflection intensities. For the reflection widths, however, the usual monotone increase of $2\theta$ was assumed.} 
To illustrate the clearly visible reflection splitting, the agreement between experimental and simulated diffractogram for a small subset of reflections is shown in Fig.\,\ref{fig:XRD-Diff}. 
The blue points indicate the experimental intensities, the orange curve represents the least-squares fit, and the red line the difference between the two. Since clearly separated, the reflex positions could be sufficiently described without additional assumptions. The resulting lattice parameters are $a=2.850$\,\AA{} and $c=2.993$\,\AA{} so that the tetragonality is given by $c/a=1.050$.  
If the measured carbon content from table \,\ref{tab:Composition} is used into equ.\,\eqref{equ:c-a}, a slightly higher tetragonality of $c/a=1.054$ is obtained. 
\mrk{
If one compares X-ray diffraction analyses with $c/a$ ratios derived from EBSD patterns, one obtains slightly lower $c/a$ for EBSD analyses \cite{tanaka2020amat}. This systematic offset was attributed to the different information depths of both methods, since the information depth relevant to EBSD is significantly lower than for XRD.
}
\mrk{Since the samples were heat-treated at 1000\degree{}C for 30\,min, the remaining cementite should be completely dissolved so that the $c/a$ reduction derived from EBSD patterns should hardly be due to an effectively lower carbon concentration in the austenite. 
}


\subsection{Microstructure characterization by EBSD}


\begin{figure}[tbh]
\begin{center}
\begin{subfigure}{8cm}
    \includegraphics[width=\textwidth]{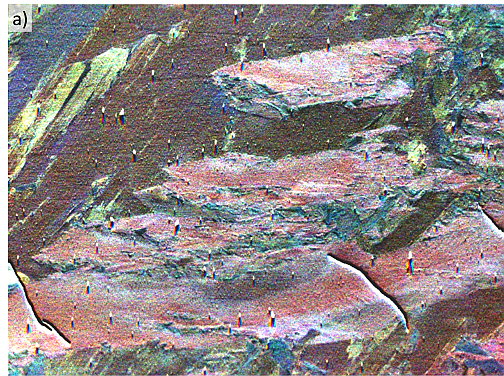}
    \subcaption{
    }
\vspace{2ex}
\end{subfigure}
\\
\begin{subfigure}{8cm}
    \includegraphics[width=\textwidth]{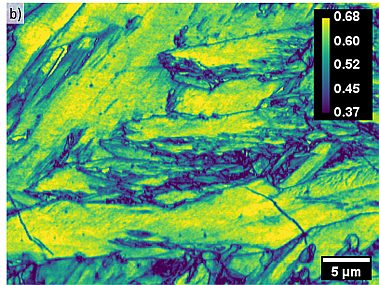}
    \subcaption{
    }
\end{subfigure}
\end{center}
\vspace{-3ex}
\caption{\label{fig:FSE} Forescattered electron (FSE) detector image (Argus, Bruker) of the mapped area (a). 
In (b) the normalized cross-correlation coefficient $r$ between two experimental Kikuchi patterns collected at adjacent horizontal positions are shown. (map size: $38\times 28$\textmu m, step size: 94\,nm) }
\end{figure}

\subsubsection{Qualitative imaging by pattern correlation}

For EBSD mappings, the normalized cross-correlation coefficient between Ki\-ku\-chi patterns collected at adjacent map positions can be used to qualitatively indicate local changes in a microstructure without even having to know the phase and crystal orientation.
As long as orientation, phase and defect density do not change significantly from measuring point to measuring point, $r$ is relatively high:\@ $0.5<r<0.7$. 
However, if one of them changes, $r$ drops to low values like $r<0.2$. 
When comparing Kikuchi patterns from adjacent map positions, $r$ maps mainly display high and low angle grain boundaries, phase boundaries, but also other lattice defects like microscopic deformation zones around scratches. 

For martensite, $r$ maps are useful to visualize microstructural features which can be difficult to interpret in SEM images.   
\mrk{Fig.\,\ref{fig:FSE} (a) represents the signal mixed as RGB colours from three independent BSE detectors mounted below the phosphor screen. The resulting image mainly shows the orientation-dependent influence of the electron channeling-in on the background of the channeling-out EBSD pattern. 
It has been shown that channeling-in is sensitive to small orientation changes  \cite{Nolze:17,Day:93,Nolze:14b,Britton2018}. }  
\mrk{Fig.\,\ref{fig:FSE} (b) displays the corresponding $r$-map which shows in dark color the grain or phase boundaries, whereas bright regions indicate single crystalline areas (grains). 
Crystal damages like the visible fissures intersecting a larger martensite region in the lower half of the image appear visible as dark lines similar to boundaries.
The same would happen due to preparation scratches, but at such high magnifications these would be visible as straight lines. 
The sample preparation in our case was optimized until such damages are no longer present or only to a very small extent, so that conclusions on the local structure of the material are not affected. 
Within the large grain in the upper left corner of (a) there are some slightly sloping linear grooves, which are most likely the last remains of grinding and polishing scratches. 
The absence of corresponding traces in the map of the neighbor cross-correlation coefficient in Fig.\,\ref{fig:FSE}(b) prove that the linear features in Fig.\,\ref{fig:FSE} (a) probably are unproblematic topography effects. 
The deformation zones around the scratches have obviously already been removed during the etching process during silica polishing. 
The sensitivity of the $r$-map is also shown by the isolated silica particles in Fig.\,\ref{fig:FSE}(a) -- cf.\@ the opposite shadows compared to the fissures -- which attenuate the diffraction signal despite their low mean atomic number, so that the corresponding map positions of these particles in Fig.\,\ref{fig:FSE}(b) appear darker.
In order to illustrate the importance of careful sample preparation, we show two examples of maps with dissatisfactory sample preprartion in the supplementary data, see Figures S1 and S2 in the appendix.
}
\mrk{In terms of microstructure, the scanned field in Fig.\,\ref{fig:FSE} contains areas of larger martensite laths or plates that are most likely almost parallel to the sample surface. 
However, there are also areas of martensite grains where the axes of the laths are inclined towards the surface so that their cross section appears much smaller. }

\FloatBarrier
\subsubsection{Matching of discrete $c/a$ ratios}

\begin{figure}[tb!]
\begin{center}
  \begin{subfigure}{0.45\textwidth}
    \includegraphics[width=\textwidth,trim=0cm 0cm 0cm 0cm, clip=true]{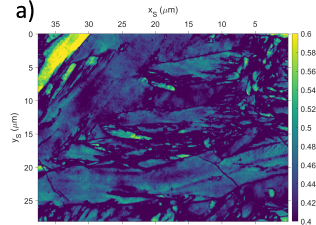}
    \caption{}
    \label{fig:c/a=1.00}
  \end{subfigure}
  \hfill
  \begin{subfigure}{0.45\textwidth}
    \includegraphics[width=\textwidth,trim=0cm 0cm 0cm 0cm, clip=true]{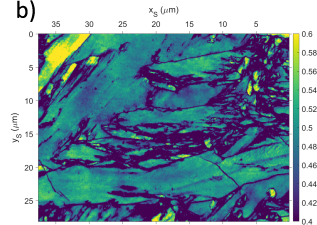}
    \caption{}
    \label{fig:c/a=1.02}
  \end{subfigure}
  \\
  \begin{subfigure}{0.45\textwidth}
    \includegraphics[width=\textwidth,trim=0cm 0cm 0cm 0cm, clip=true]{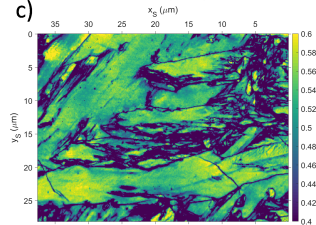}
    \caption{}
    \label{fig:c/a=1.04}
  \end{subfigure}
  \hfill
  \begin{subfigure}{0.45\textwidth}
    \includegraphics[width=\textwidth,trim=0cm 0cm 0cm 0cm, clip=true]{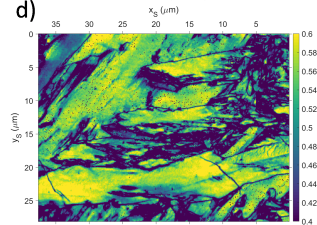}
    \caption{}
    \label{fig:c/a=1.06}
  \end{subfigure}
  \\
  \begin{subfigure}{0.45\textwidth}
    \includegraphics[width=\textwidth,trim=0cm 0cm 0cm 0cm, clip=true]{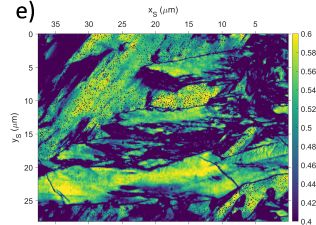}
    \caption{}
    \label{fig:c/a=1.08}
  \end{subfigure}
  \hfill
  \begin{subfigure}{0.45\textwidth}
    \includegraphics[width=\textwidth,trim=0cm 0cm 0cm 0cm, clip=true]{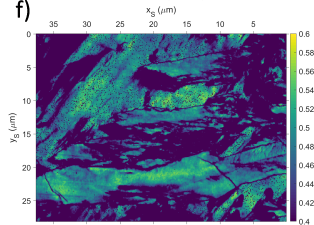}
    \caption{}
    \label{fig:c/a=1.10}
  \end{subfigure}
    \vspace{-3ex}
\end{center}
\caption{\label{fig:r-c/a} $r$-maps for master patterns of martensites assuming different $c/a$ ratios: (a) $c/a=1.00$, (b) $c/a=1.02$, (c) $c/a=1.04$, (d) $c/a=1.06$, (e) $c/a=1.08$, (f) $c/a=1.10$.}
\end{figure}

In Fig.\,\ref{fig:r-c/a} we show the results of pattern matching analyses for six hypothetical martensites of different tetragonality. The template patterns were calculated for $c/a = 1.00, 1.02, 1.04\ldots 1.10$.

For the more blurred Kikuchi patterns of martensite, NCC values are typically in the range $0.4\leq r \leq 0.6$.
The various $r$ maps show microstructure areas where the $r$ is quite high, while other areas consistently show lower $r$. 
It is also noticeable that the maximum $r$ values appear at different areas in the map depending on $c/a$, see Fig.\,\ref{fig:r-c/a}. 
This observation suggest a high probability of locally different $c/a$ ratios in martensite since neither the pattern quality map (not shown) nor the cross-correlation between patterns from adjacent measurement positions in Fig. \,\ref{fig:FSE}(b) show structures that explain the observed variations in $r$ values by an effect other than a local variation in $c/a$ values.
In accordance with the result from the XRD measurements, the $r$ map in Fig.\,\ref{fig:c/a=1.06} for $c/a=1.06$ displays the highest percentage of measurement points with maximum $r$, i.e.\@ the largest amount of bright yellow pixels.

\begin{figure}[ht]
\centering
    \includegraphics[width=0.6\columnwidth,trim=.0cm 0.05cm 0.05cm 0.70cm, clip=true]{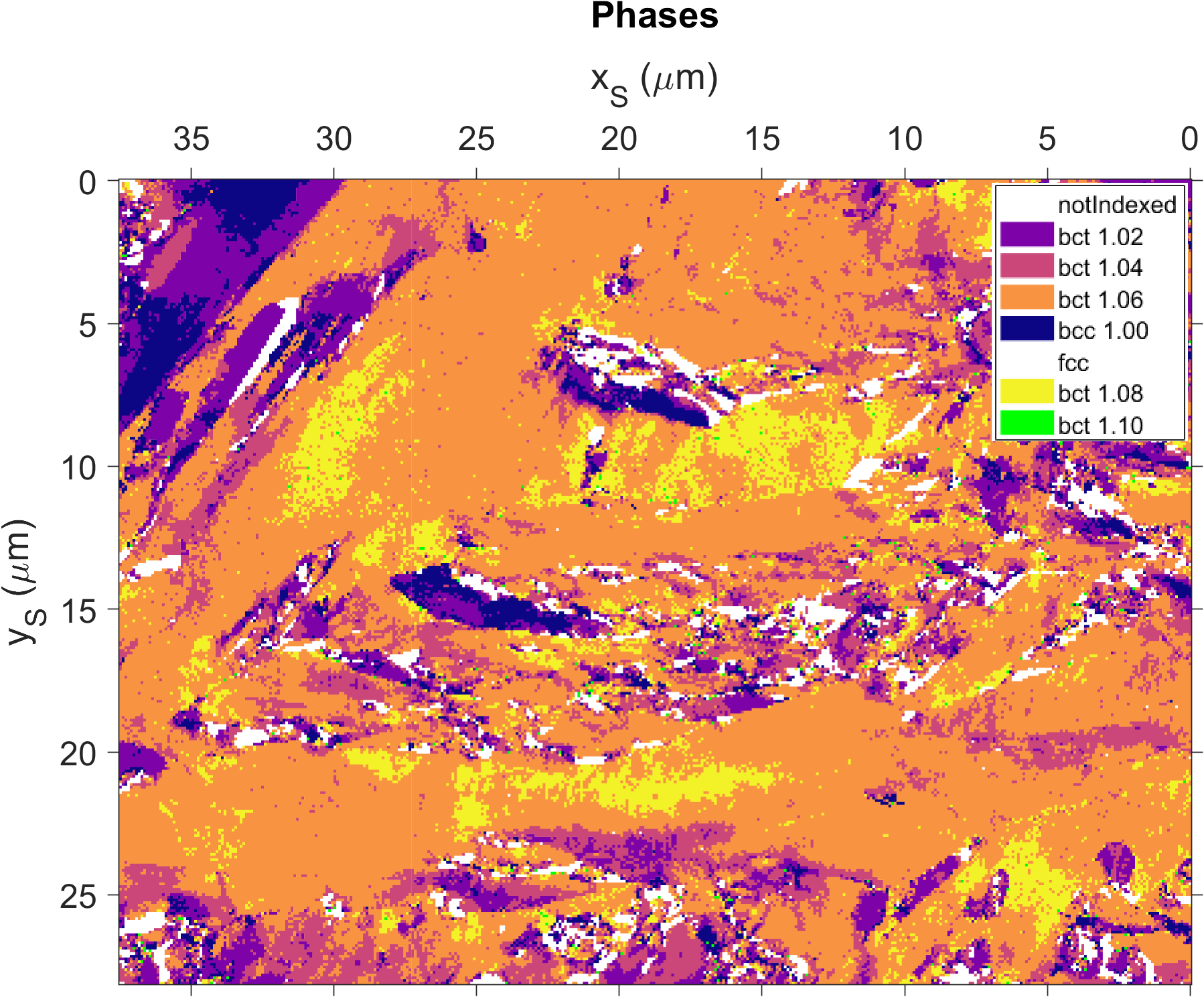}
    \caption{Map of the best-fit $c/a$ with respect to the maximum cross-correlation coefficient $r$. ($c/a=1\ldots 1.1$, $\Delta(c/a)=0.02$) }
    \label{fig:manual_c-to-a}
\end{figure}

In order to show the most probable $c/a$ for each local Kikuchi pattern, i.e. for each measured map point, a specific color was assigned to each tested $c/a$ ratio. For each map point,  Fig.\,\ref{fig:manual_c-to-a} shows the $c/a$-color for which the highest $r$ value, i.e. the best fit, was obtained. 
White areas indicate retained austenite on the map.

The following conclusions can be drawn from the results discussed so far and summarized in Fig.\,\ref{fig:manual_c-to-a}:  
\begin{enumerate}[a)]
\setlength\itemsep{1em}
\item By pattern matching, even blurred Kikuchi patterns of martensite can be identified as systematically tetragonally distorted. 

\item If we take the maximum correlation coefficient $r$ as a criterion for the locally most probable $c/a$, the resulting $c/a$-map shows a complex distribution of areas with different $c/a$.  

\item Most of the area studied provides a maximum correlation for $c/a=1.06$. This is comparatively close to the XRD result ($c/a=1.050$) and also to the theoretical prediction ($c/a=1.055$) derived from the assumed carbon content. 

\item The local variation of $c/a$ determined by EBSD could explain the reflex broadening in X-ray diffractograms typical of martensite \cite{Seljakow:27,Kurdjumov:76}.

\end{enumerate}

\begin{figure}
\begin{center}
 \begin{subfigure}{0.6\textwidth}
\includegraphics[width=\textwidth,trim=0cm 0cm 0cm 0.0cm, clip=true]{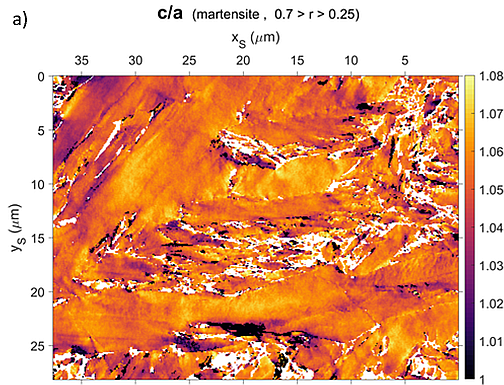}
  \caption{}
  \label{fig:bct-PM}
 \end{subfigure}
 \vspace{12pt}
 \\
    \begin{subfigure}{0.6\textwidth}
    \centering
    \includegraphics[width=0.75\textwidth,trim=0cm 0cm 0cm 0.0cm, clip=true]{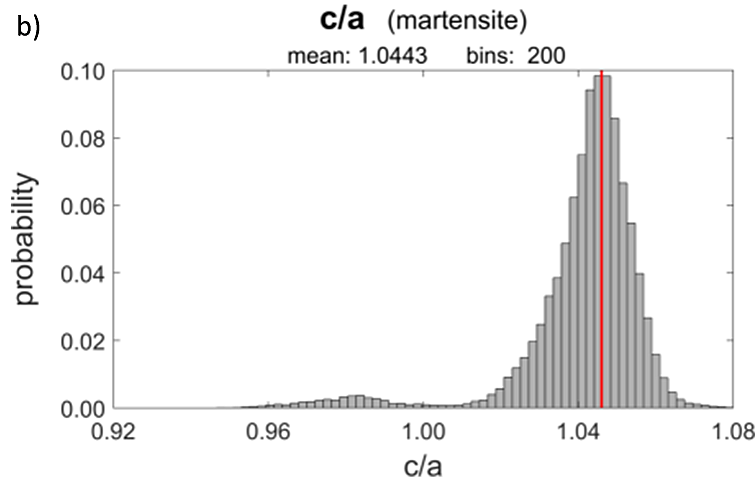}
    \subcaption{}
    \label{fig:bct-PMhist}
    \end{subfigure}
 \vspace{12pt}
 \\
    \begin{subfigure}{0.6\textwidth}
\includegraphics[width=\textwidth,trim=0cm 0cm 0cm 0.0cm, clip=true]{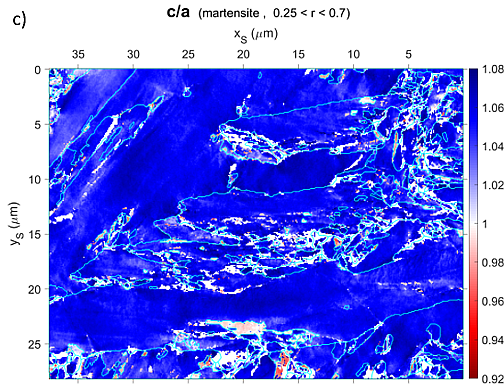}
  \caption{}
  \label{fig:bct-PM-BR}
 \end{subfigure}
\vspace{-3ex}
\end{center}
\caption{Lattice-variable pattern matching on martensite, allowing for a continuously variable $c/a$ ratio. The histogram (b) also indicates a small amount of grains which are characterized by $c/a<1$. The respective grains can be seen in red in (c) which displays (a) with a modified color bar covering $c/a\lessgtr 1$.}
\label{fig:variablePM}
\end{figure}

\subsubsection{Lattice-variable pattern matching}

As a second approach to evaluate the inherent tetragonality hidden in experimental EBSD patterns, we have applied a projective transformation approach \cite{Winkelmann:18}.
\mrk{This approach is based on the influence of the lattice parameter \emph{ratios} on the Kikuchi pattern, which leads to a recognizable shift of the zone axes, even in highly binned diffraction patterns. In contrast, due to the small Bragg angles and the correspondingly small number of pixels within a Kikuchi band profile, the same change in $c/a$ has a negligible influence on the bandwidth variation resulting from the applied transformation \cite{Nolze:17}.}

\mrk{Template patterns are reprojected in real time for varying lattice parameters by applying an appropriate image transformation to a fixed reference simulation. 
In this way, a time-consuming re-simulation can be avoided for a non-cubic phase with varying lattice parameters, e.g. solid solutions, strained microstructures, gradient materials with noticeable diffusion gradients, etc.
A comparison of exemplary Kikuchi patterns for different tetragonal distortions is given in \cite{Winkelmann:18}, together with an analysis of the effect of the projective distortion relative to simulations with fixed $c/a$ as discussed above. 
The analysis in \cite{Winkelmann:18} indicated that the sensitivity of the projective pattern matching approach is about 0.5 percent change in $c/a$ for an assumed pattern resolution of $100\times 100$ pixels.
}

The result of this lattice-variable pattern matching is shown in Fig.\,\ref{fig:variablePM}a. 
Despite the $160\times 115$ pixel size of the applied Kikuchi patterns, the greater variability of the fitting means that significantly more microstructural details are visible. 
With each independently analyzed Kikuchi pattern the smooth variation of $c/a$ in Fig.\,\ref{fig:variablePM}a confirms a local variation of the lattice parameters in the investigated martensitic microstructure \cite{Seljakow:27}.
These variations are consistent with the typical reflex broadening for martensite in XRD diffractograms, as displayed exemplarily in Fig.\,\ref{fig:XRD-Diff}.

\begin{figure}[htb]
    \centering
    \includegraphics[width=0.7\textwidth,trim=0cm 0cm 0.0cm 0.6cm, clip=true]{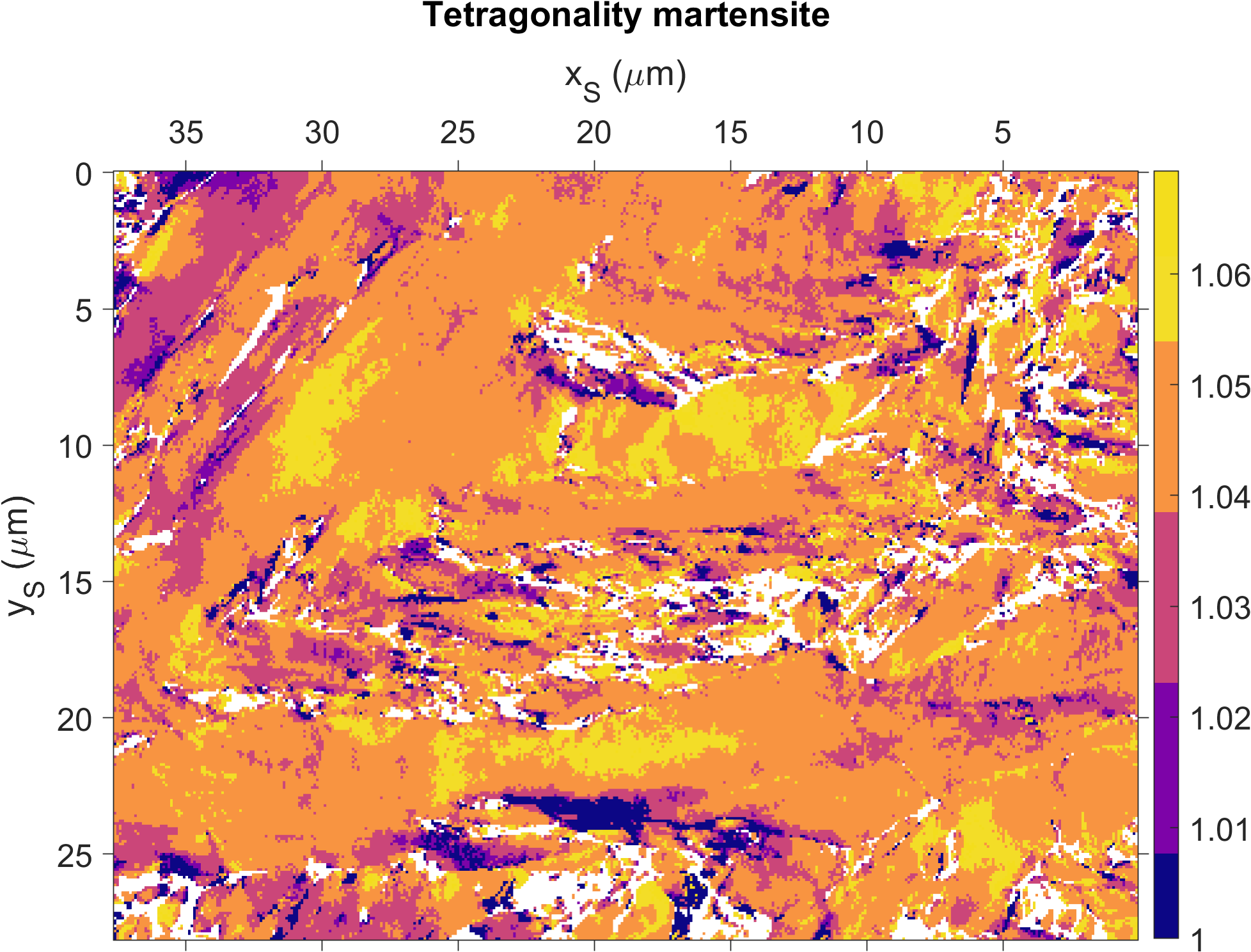}
    \caption{Image with discrete $c/a$ levels derived from the pattern matching analysis displayed in Fig.\,\ref{fig:bct-PM}, for comparison to Fig.\,\ref{fig:manual_c-to-a}. The white areas indicate austenite and locations where the cross correlation coefficient $r$ was below 0.25.}
    \label{fig:my_label}
\end{figure}

\mrk{To compare the results of the two EBSD analysis approaches, which are based on fitting a set of discrete $c/a$ ratios vs. continuously varying $c/a$ ratios, the map in Fig.\,\ref{fig:bct-PM} was discretized to the same color levels as in Fig.\,\ref{fig:manual_c-to-a}. 
The result in Fig.\,\ref{fig:my_label} is convincingly similar to the map shown in Fig.\,\ref{fig:manual_c-to-a}, i.e. most of the $c/a$ values show a good match in both maps. 
This illustrates the effectiveness of the projective transformation approach based on a single reference pattern simulation, e.g. for $c/a=1$, with the derivation of the template patterns for varying $c/a$ by a corresponding projection of this single reference pattern.}

Similar results have recently been presented in \cite{Tanaka:18}, but using Kikuchi patterns with maximum image resolution. 
The variation in derived tetragonality covered a range from $c/a=0.98$ to $1.08$, although due to the lower carbon content of only 0.77\,m\%\,C the expected tetragonality is lower at $c/a=1.035$. From the measurements it was concluded that the martensitic structure is composed of grains with different $c/a$ ratios, any variations of $c/a$ within individual grains were not discussed. However, both interpretations lead to the same conclusion: The observed reflex broadening in the martensite is at least partly the result of locally varying lattice parameters.

\mrk{
The $c/a$-distribution of the scanned area in Fig.\,\ref{fig:bct-PM} reflects a mean value of $c/a=1.044$, i.e.\@ slightly below the X-ray result found at 1.050.
In agreement with this observation, in \cite{tanaka2020amat} slightly lower $c/a$ values were derived from high-resolution EBSD patterns compared to XRD measurements. The observed offset was attributed to the different information depths of XRD and EBSD because the data measured by EBSD are representative of a much lower information depth.
}

\subsubsection{Meaning of fit results with $c/a<1$}

In the pattern matching analysis with discrete tetragonality, we did not explicitly consider simulations for $c/a<1$, while the lattice variable pattern matching did not have this limitation.
The histogram in Fig.\,\ref{fig:bct-PMhist} indicates that a small amount of martensite characterized by $c/a<1$ is present in the microstructure shown. 
The alternative coloration of the martensite tetragonality map in Fig.\,\ref{fig:bct-PM-BR} shows that the peak at about $c/a=0.985$ in the histogram is formed by regions in the map which are shown in reddish colors.

While $c/a<1$ could indicate a shortened $c$-axis of the tetragonal unit cell, such a shortening would not be expected from previous observations of martensite tetragonality \cite{cheng1990smet}.
Instead, the fitted lattice parameter \emph{ratio} of $(c/a)_\textrm{fit} < 1$ is still consistent with an expansion of the martensite lattice, but it points to the role of a possible simultaneous lattice expansion along two orthogonal directions. This can be seen by the following argument. 
Since we use a fit procedure which varies the lattice distortion along a single axis only, a possible way for the fit algorithm to nevertheless achieve a fit to two elongated axes is to rotate the former $c$ axis to the $a$-direction and fit the inverse ratio $(a/c)_\textrm{fit}$ with a value \emph{smaller} than 1.0. 
Effectively, this treats \emph{two} \emph{increased} lattice parameter ratios $c/a$ and $b/a$.
A related observation has been reported in \cite{tanaka2020amat}, where some fit results on martensite samples are interpreted by the presence of an orthorhombic distortion. 
\textsc{Kurdjumov} and \textsc{Khachaturyan} \cite{kurdjumov1975amet,Kurdjumov:76} also discussed further peculiarities of tetragonal and orthorhombic martensite in steels. For example, in Figure 2a of \cite{kurdjumov1975amet}, two lattice parameters have been observed with different values of similar size, while the third lattice parameter is smaller.
We thus conclude tentatively that the map areas with tetragonality values $(c/a)_{fit} < 1$  actually correspond to the inverse value of $(a/c)_\textrm{fit} > 1$ and an additional elongated $b$-axis.
Effectively, the tetragonality fit approximated the general orthogonal distortion by a pseudosymmetric tetragonal setting of the unit cell, which has the shortest orthorhombic axis along a contracted tetragonal axis.
The implementation and investigation of the effects of a general orthorhombic distortion in the fit of lattice parameter ratios will be the subject of future studies.

\FloatBarrier
\subsubsection{Calibration: Lattice-variable pattern matching on austenite}

\begin{figure}[th]
  \centering
   \begin{subfigure}{0.6\textwidth}
     \includegraphics[width=\textwidth,trim=0cm 0cm 0.0cm 0.0cm, clip=true]{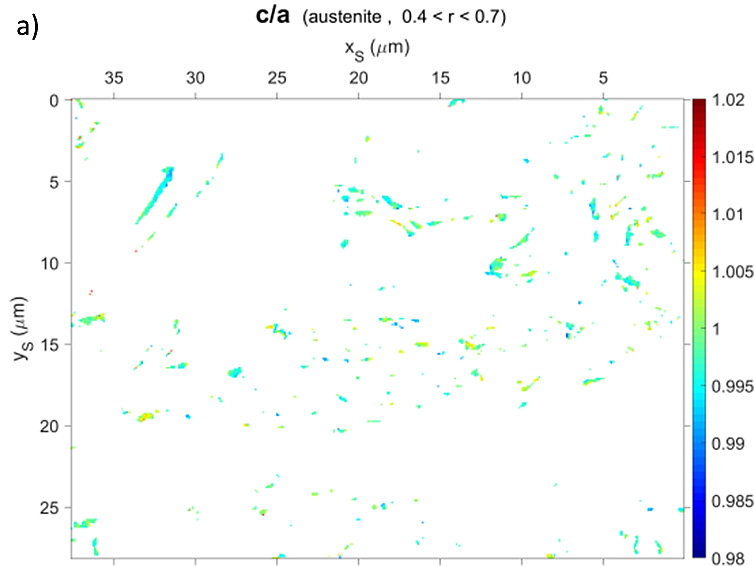}
     \caption{}
     \label{fig:fcc-PM}
   \end{subfigure}
   \vspace{6pt}
\\
   \begin{subfigure}{0.6\textwidth}
     \includegraphics[width=\textwidth,trim=0cm 0cm -0.6cm 0.0cm, clip=true]{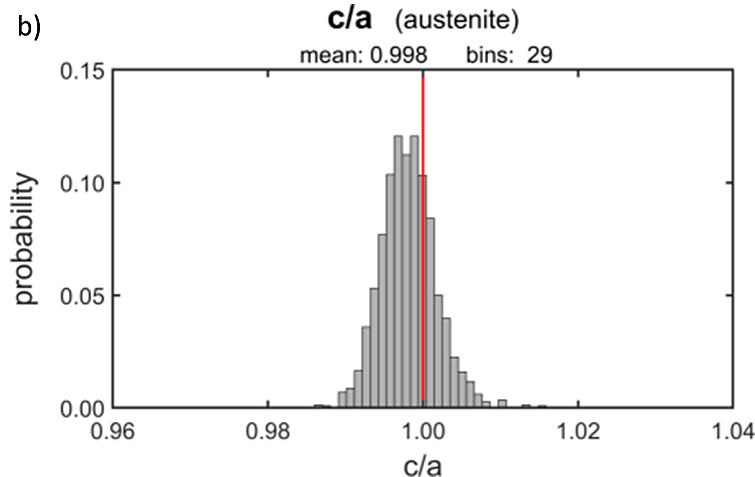}
     \subcaption{}
     \label{fig:fcc-ca}
   \end{subfigure}
\caption{Variable-lattice pattern matching of austenite, (a) c/a map, (b) probability histogram. All data displayed have a higher cross-correlation coefficient than 0.4.}
\label{fig:Histogram_tetra}
\end{figure}

In order to check the plausibility of the variable-lattice pattern matching, the retained austenite was also analyzed with regard to a hypothetical tetragonal distortion. 
Assuming that the diffraction geometry (projection center) was calibrated by assuming a cubic austenite reference, we would expect a resulting lattice parameter ratio equal to one. 
This was confirmed by the lattice variable pattern matching. The $c/a$ histogram for the retained austenite in Fig.\,\ref{fig:fcc-ca} shows a peak with $0.2\%$ deviation from the ideal value $c/a=1.0$. 
The local distribution of austenite in the microstructure and the derived $c/a$ are shown in the map of Fig.\,\ref{fig:fcc-PM}. Note the much smaller color bar range ($\pm2\%$). 

We observe that the $c/a$-histogram of the austenite shows a maximum shifted slightly from the ideal position at about 0.998.
We attribute this shift to a systematic influence of the excess-deficiency effect \cite{winkelmann2008um} when adjusting the tetragonal distortion or determining the position of the projection center. 
Compared to the tetragonality found in martensite, however, the extent of this shift is marginal. 
The peak width of the histogram in Fig.\,\ref{fig:Histogram_tetra} is characterized by a standard deviation of 0.01, which can also be used as an estimate of the precision of the martensite tetragonality. 
This estimate is consistent with the martensite tetragonality distribution as shown in Figure \ref{fig:bct-PMhist}(a). 
The observed peak width of the tetragonality for austenite in Fig.\ref{fig:Histogram_tetra}(a) is limited by the pattern resolution we used in this study ($160 \times 115$ pixels) and by the degree of correlation between orientation and tetragonality during the simultaneous fitting of both parameters. 

\FloatBarrier
\subsubsection{Discussion of the observed martensite tetragonality}

\mrk{
The observations which were presented above prove a considerable variation of tetragonality within the microstructure of the high carbon steel which we investigated. 
Because we can assume that the carbon concentration in the austenite before the martensitic transformation is nearly homogeneous, the variable tetragonality in the final martensite points to a combined influence of local strains, which might influence the carbon diffusion during the long-term ageing process.
Our observations thus do not seem to be compatible with random fluctuations of unit cell distortions which were suggested to lead to an effective tetragonality in \cite{lobodyuk2018mmta}.
Instead, our observations indicate the local presence of a true tetragonal or orthorhombic structure, with changing lattice parameter ratios at different locations in dependence on local strain \cite{maugis2020jped} and possibly orientation.
Moreover, because of the displacive, non-diffusional, nature of the martensitic transformation, the positions of the carbon atoms immediately after the transformation have to be consistent with models which where suggested for the actual pathways of the Fe atoms during the transformation from austenite \cite{jaswon1948ac,cayron2018crystals}.  While we did not observe the situation immediately after the transformation in the present investigation, these models imply boundary conditions at the onset of further carbon diffusion in the presence of local strains.
Further higher-resolution EBSD investigations of tetragonal martensite  in the presence of retained austenite are expected to shed more light on the role of the local strains. 
Beyond the application to tetragonal martensite, we expect that the method presented here will also be useful for investigation of tetragonality in ferrites in general \cite{bhadeshia2013pm,pereloma2016mst}, or any other gradient material with variable lattice parameter ratios.
}

\section{Summary and Conclusions}

We investigated the tetragonality of martensite in high carbon steel (1.2\,m\% C) using X-ray diffraction and EBSD, and we compared two approaches for the determination of the local lattice parameter ratios by pattern matching to simulated EBSD reference data. 
The first approach used a discrete set of exactly simulated reference patterns for fixed $c/a$ values, while the second approach allowed to approximate the effects of lattice distortions by distortion of a fixed reference patterns to generate in real-time template patterns of sufficient quality for variable $c/a$ \cite{Winkelmann:18}. 

We find that the results of both EBSD approaches closely agree, confirming a continuous variation of $c/a$ within the investigated martensite, within a range of  $c/a$ that is consistent with the XRD results.
The lattice-variable pattern matching approach has the significant advantage that no additional dynamical EBSD simulations are necessary to approximate small changes in lattice parameter ratios. 
This makes it very efficient for the routine analysis of large EBSD data sets, as has been demonstrated for orientation maps with $400\times 300$ points in the present study.

The observed locally varying $c/a$ ratio in martensite indicates that an observed peak broadening in XRD is not necessarily the result of inherent lattice defects but can be also caused by a local modulation of the lattice parameters as already pointed out in \cite{Seljakow:27}. 
Such variability of lattice parameters can have its reason in an inhomogeneous distribution of carbon, or, for example, by the formation of a complex local strain state caused by the transformation stress of three-dimensionally intertwined variants of martensite. 

In the present study, we have demonstrated that  standard resolution Kikuchi patterns ($160\times 115$ pixels) ($\approx20k$) can be used to investigate the local tetragonality of martensite with a precision of about $0.01$ in $c/a$. 
We expect that higher resolution patterns will allow us to obtain a more detailed picture on the local lattice distortions in martensite samples, including the presence of locally orthorhombic lattice parameters which has been indicated by the current results.

\section*{Acknowledgments}

We are grateful to Ralf Hielscher (TU Chemnitz) for help with MTEX, which was used to generate all EBSD-related images. We also thank Romeo Saliwan Neumann and Michaela Buchheim (BAM) for their invaluable help during sample preparation and measurements.
This work has been supported by the Polish National Agency for Academic Exchange (NAWA) grants \\ no. PPI/APM/2018/1/00049/U/001 \\and no. PPN/ULM/2019/1/00068/U/00001.


\section*{Appendix: Effects of Sample Preparation} \label{sec:sampleprep}

\begin{figure}[!htbp]
    \renewcommand{\thefigure}{A1}
    \centering
    \includegraphics[width=0.8\columnwidth,trim=0cm 0cm 0.0cm .0cm, clip=true]{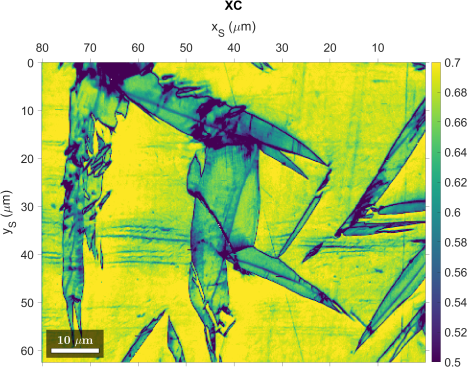} \\
    \vspace{0.8cm}
    \includegraphics[width=0.8\columnwidth,trim=0cm 0cm 0.0cm .0cm, clip=true]{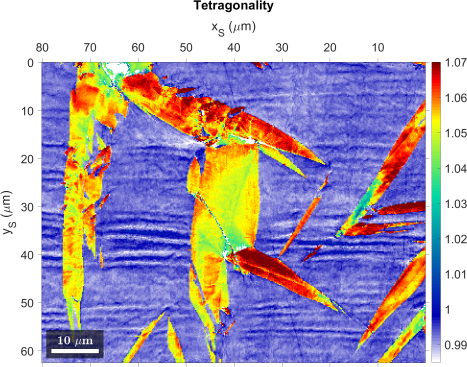}
    \caption{ Top: Normalized Cross Correlation Coefficient (XC) of the experimental patterns relative to the best-fit simulations.
    Bottom: Apparent tetragonality variations in the austenite (blue). The observed tetragonality $c/a$ in the martensite is significantly larger (yellow/red) and shows tetragonality features which are not correlated with the scratch/deformation features.}
    \label{fig:prep1}
\end{figure}

\begin{figure}[!htbp]
\renewcommand{\thefigure}{A2}
    \centering
    \includegraphics[width=0.8\columnwidth,trim=0cm 0cm 0.0cm .0cm, clip=true]{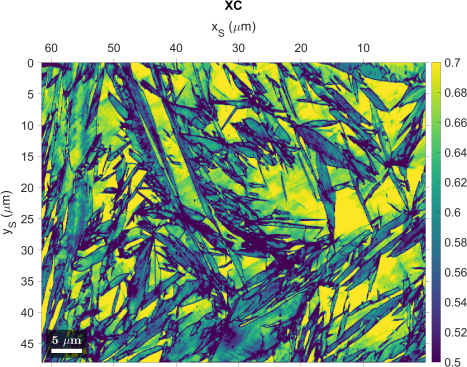} \\
    \vspace{0.8cm}
    \includegraphics[width=0.8\columnwidth,trim=0cm 0cm 0.0cm .0cm, clip=true]{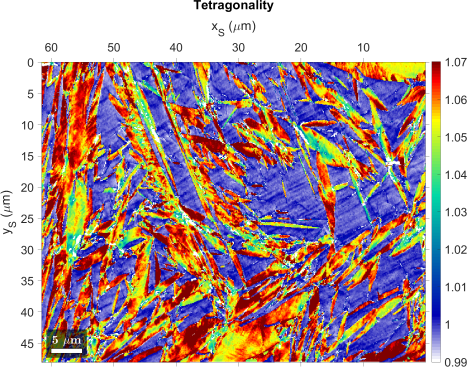}
    \caption{Top: Normalized Cross Correlation Coefficient (XC) of the experimental patterns relative to the best-fit simulations.
    Bottom: Apparent tetragonality variations in the austenite (blue). The observed tetragonality $c/a$ in the martensite is significantly larger (yellow/red) and shows tetragonality features which are not correlated with the scratch/deformation features.}
    \label{fig:prep2}
\end{figure}

In order to emphasize the necessity of very careful sample preparation for measurements of changes in lattice parameter ratios by EBSD,  we show in Fig.\,\ref{fig:prep1} and Fig.\,\ref{fig:prep2} example maps which resulted from a dissatisfactory sample preparation. 
The maps were not considered for the quantitiative analysis as presented in the main text because of preparation artefacts which became visible during the pattern matching analysis. 
The preparation artefacts can be seen as nearly parallel scratches predominantly in the austenite regions with a resulting $c/a$ varying systematically near 1.0.
Some of the visible features in the austenite near the grain boundaries of martensite are possibly due to the relaxation of phase transformation stresses during the sample preparation, compare e.g. Figure 1 in \cite{bowles1952pimp}.
While some features in the measurements of Fig.\,\ref{fig:prep1} and Fig.\,\ref{fig:prep2} cannot be conclusively interpreted, these maps demonstrate the high sensitivity of the pattern matching approach to systematic changes in the experimental Kikuchi patterns, which can be caused by various effects.

\FloatBarrier

\end{document}